\documentstyle[12pt,amssymb]{article}

\def\be{\begin{equation}}
\def\ee{\end{equation}}

\title{
Hypergeometric solutions of some algebraic equations\,\footnote{\,
This paper aroused from the problem of normalization of the ground state
wave function of the trigonometric $n$-particle Calogero-Sutherland system
(see the review paper [OP 1983]). The relation of this problem with the
results of the present note will be considered later.}}

\author{A.M. Perelomov
\footnote{On leave of absence from Institute for Theoretical and Experimental
Physics, 117259 Moscow, Russia. Current e-mail address:
perelomo@dftuz.unizar.es}\\
{\small\em
Max-Planck-Institut~f\"ur~Mathematik,}\\
{\small\em Vivatsgasse 7, D-53111 Bonn, Germany }}

\date{}
\begin{document}
\maketitle{}

\centerline{To the memory of A.N. Tyurin}

\begin{abstract}\noindent
We give the hypergeometric solutions of some algebraic equations including
the general fifth degree equation.
\end{abstract}

\setcounter{equation}{0}

\medskip\noindent {\bf 1.}
It is well known that the general algebraic equation of degree $n\geq 5$
cannot be solved by radicales [Ab 1826]. However, it may be solved if
we use  wider classes of functions.
For example, the fifth degree equation may be solved in modular functions
[He 1858], [Kr 1858], [Kl 1884]; the general algebraic equation
may be solved in hyperelliptic theta constants [Um 1984].

Another approach consists in the consideration of the algebraic solutions
of the differential equations and was used in the classical Schwarz's paper
[Sch 1873], where all algebraic solutions of the standard hypergeometric
equation were found. It was intensively developed by Poincar\'e.

The results of Schwarz were completed in the paper [BH 1989] by the
classification of the algebraic generalized hypergeometric functions
$\,_nF_{n-1}$.
There is one series corresponding to the case under consideration, and the
76 exceptional cases. Note that for $n\geq 9$ there are no  exceptional
cases.

In [BH 1989] the monodromy groups for these functions were calculated
explicitly. They are the groups generated by the complex reflections in
${\Bbb C}^n$. The differential Galois groups of the corresponding
hypergeometric differential equations are also explicitly calculated there.
However, the corresponding algebraic equations were not calculated.

In this note we give the explicit solution of the equations of type
\begin{eqnarray}
F_n(x,t) &\equiv & x^n-x+t =0,\qquad n>1, \\
G_n(y,t) &\equiv & \gamma _n t^{n-1}y^n-(y-1)\left( y+\frac1{n-1}\right)
^{n-1}=0,\qquad \gamma _n=\frac{n^n}{(n-1)^{n-1}}\nonumber \\
&& \end{eqnarray}
in the class of hypergeometric functions of type
\be _{n-1}F_{n-2}(\alpha _1,\ldots ,\alpha _{n-1};\,\beta _1,\ldots ,
\beta _{n-2};\,\gamma t^{n-1}).\ee
Note that the general equations of degrees  2,\,3,\,4 and 5  
may be reduced to equation (1) by Tschirnhaus transformation, 
see [Fr 1924]\,\footnote{\,After this work had been completed,
I learned from Prof. Zagier that the solution of equation (1) is also
known to him, unpublished.}.

\medskip\noindent{\bf 2.}
First, let us remark that the normalization of the ground state wave
function of the $n$-particle Calogero-Sutherland problem is equivalent to
the calculation of the integral
\be \int _0^{\pi }...\int _0^{\pi } \left\{ \prod _{1\leq j<l\leq n}
\sin (q_j-q_l)\right\} ^k dq_1 \cdots dq_n. \ee
In the process, there appear binomial coefficients of special type
\be c_k^{(n,j)}=\left( \begin{array}{c} nk+j \\ k\end{array} \right)
=\frac{(nk+j)!}{((n-1)k+j)!\,k!}\,. \ee
It is natural to introduce for them the generating functions
\be H_{n,j}(z)=\sum _{k=0}^\infty c_k^{(n,j)}z^k,\qquad n\geq 2,\quad
j=0,1,\ldots . \ee
Let us introduce also the functions
\begin{eqnarray}
y_{n,j}(t) &=& t^j\,H_{n,j}(z),\qquad \mbox{where}\quad z=t^{n-1},\quad
j=0,1,\ldots . \\
x_{n,j}(t) &=& j \int _0^t y_{n,j-1}(\tau )\,d\tau ,\qquad j=1,2,\ldots \,.
\end{eqnarray}
These functions  have a number of nice properties, in particular,
$x_{n,1}(t)$ and $y_{n,0}(t)$ are the solutions of  equations (1) and (2).

Recall the definition of the generalized hypergeometric function:
\begin{eqnarray}
&&\,_mF_n(\alpha _1,\alpha _2,\ldots ,\alpha _m;\beta _1,\beta _2,\ldots
\beta _n;\gamma z) =\sum _{k=0}^\infty c_kz^k= \nonumber \\
&& \sum _{k=0}^\infty \frac{(\alpha _1)_k\,(\alpha _2)_k\cdots (\alpha _m)_k}
{(\beta _1)_k\,(\beta _2)_k\cdots (\beta _n)_k\,k!}\,(\gamma z)^k,
\end{eqnarray}
where
\[ (\alpha )_k=\alpha (\alpha +1)\cdots (\alpha +k-1)\]
is the Pochhammer symbol.

\medskip\noindent{\bf Lemma 1.} {\em The functions} (6)--(8) {\em are
algebraic and of hypergeometric type. Namely,}
\be H_{n,j}(z)=\,_{n-1}F_{n-2}\left( \frac{j+1}{n},\ldots ,\frac{j+n}{n};
\frac{j+1}{n-1},\ldots ,\frac{j+n-1}{n-1};\gamma _nz\right) , \ee
{\em where}
\be  \gamma _n=\frac{n^n}{(n-1)^{n-1}}\,.\ee
{\em The functions} $y_{n,j}(t)$ {\em are given by formula} (7). {\em We have 
also}
\begin{eqnarray}
&&x_{n,j}(t)=\nonumber \\
&&t^j\,_{n-1}F_{n-2}\left( \frac{j}{n},\frac{j+1}{n},
\ldots ,\frac{j+n-1}{n};\frac{j+1}{n-1},\frac{j+2}{n-1},\ldots ,
\frac{j+n-1}{n-1};\gamma _nz\right) .\nonumber \\
&&\end{eqnarray}

\medskip\noindent{\em Proof.} This follows from the fact that for the
generalized hypergeometric function (9) 
\be \frac{c_{k+1}}{c_k}=\frac{(k+\alpha _1)(k+\alpha _2)\cdots (k+\alpha _m)}
{(k+\beta _1)(k+\beta _2)\cdots (k+\beta _n)}\,\gamma , \ee
and we have the identities
\begin{eqnarray}
&&\,_nF_{n-1}(\alpha _1,\ldots ,\alpha _{n-1},\alpha ;\beta _1,\ldots ,
\beta _{n-2},\alpha ;z)=\nonumber \\
&&\,_{n-1}F_{n-2}(\alpha _1,\ldots ,\alpha _{n-1};\beta _1,\ldots ,
\beta _{n-2};z).\end{eqnarray}

The algebraic character of the functions $H_{n,j}(z)$ becomes evident after
the comparison of (10) with the list of functions from [BH 1989].

The hypergeometric character of these functions gives the possibility to
easily differentiate and integrate them. For this we may use the
well-known formulae:
\begin{eqnarray}
&&\frac{d}{dz}\,_mF_n(\alpha _1,\ldots ,\alpha _m;\beta _1,\ldots ,\beta _n;
z) =\nonumber \\
&&\frac{\alpha _1\cdots \alpha _m}{\beta _1\cdots \beta _n}\,_mF_n(\alpha _1
+1,\ldots ,\alpha _m+1;\beta _1+1,\ldots ,\beta _n+1;z), \end{eqnarray}
\begin{eqnarray}
&&\int \,_mF_n(\alpha _1\cdots \alpha _m;\beta _1\cdots \beta _n;z)\,dz
=\nonumber \\
&&\frac{(\beta _1-1)\cdots (\beta _n-1)}{(\alpha _1-1)\cdots (\alpha _m-1)}
\,_mF_n(\alpha _1-1,\cdots ,\alpha _m-1; \beta _1-1,\cdots ,
\beta _n-1;z) +\mbox{const}.\nonumber \\
&&\end{eqnarray}
From these formulae we obtain
\begin{eqnarray}
&&\frac{d}{dz}\,H_{n,j}(z)=\nonumber \\
&&(j+n)\,_{n-1}F_{n-2}\left( \frac{j+n+1}{n},\ldots ,
\frac{j+2n}{n};\frac{j+n}{n-1},\ldots ,\frac{j+2n-2}{n-1};\gamma _nz\right) 
\nonumber \\
&& \end{eqnarray}
and formula (12).

\medskip\noindent{\bf 3.}
Note that equation (1) has the solution $x(t)$, $x(0)=0$, analytic in the
disc $D_{r_n}$ of some radius $r_n>0$. Consider also the function
$y(t)=x'(t)$. Differentiating (1) with respect to $t$, we obtain
\be \left( 1-nx^{n-1}\right) y=1. \ee
Equation (1) yields also
\be x^{n-1} =\frac1{n}\left( 1-\frac1{y}\right) . \ee
From this it follows equation (2).

The main result of this paper is 

\medskip\noindent{\bf Theorem 1.} {\em The solutions of equations} (1)
{\em and} (2) {\em are given by the formulae}
\be x(t)=x_{n,1}(t),\qquad y(t)=y_{n,0}(t). \ee
{\em We also have}
\be (x(t))^j =x_{n,j}(t),\qquad (x(t))^j\,y(t)=y_{n,j}(t). \ee

\medskip\noindent{\bf 4.} Before to prove this theorem, we  give some 
properties of generalized hypergeometric functions (for more details see 
[Po 1888] and [BE 1953]).

The generalized hypergeometric function $\,_nF_{n-1}(z)$ is defined by the
series
\be _nF_{n-1}(\alpha _1,\ldots ,\alpha _n; \beta _1,\ldots ,\beta _{n-1};z)
=\sum _{k=0}^\infty \frac{(\alpha _1)_k\cdots (\alpha _n)_k}
{(\beta _1)_k\cdots (\beta _{n-1})_k\,k!}\,z^k \ee
which converges in the unit disc.

The function $u(z)=\,_nF_{n-1}(z)$ satisfies the hypergeometric
equation \be D\cdot (D+\beta _1-1)\cdots (D+\beta
_{n-1}-1)u-z\,(D+\alpha _1)\cdots (D+\alpha _n)\,u=0 \ee with
$D=z\,{d}/{dz}$. This function is analytic on the Riemann sphere
${\Bbb C}P^1={\Bbb C}^1\cup \{\infty \}$ except the points $z=0$,
$z=\infty $ and $z=1$. The local behaviour of this function at
these points is defined by the exponents \be \begin{array}{ll}
1-\beta _1,\ldots ,1-\beta _{n-1},0 & \mbox{at}\quad z=0,\\
\alpha _1,\ldots ,\alpha _n & \mbox{at}\quad z=\infty ,\\
0,1,\ldots ,n-2,\quad \gamma =\sum _{j=1}^{n-1} \beta _j-
\sum _{l=1}^n \alpha _l & \mbox{at}\quad z=1. \end{array} \ee
If the numbers $\beta _1,\ldots ,\beta _{n-1}$ are distinct mod$\,{\Bbb Z}$,
then $n$ independent solutions of equation (21) are given by the formula
\be z^{1-\beta _i}\,_nF_{n-1}(1+\alpha _1-\beta _i,\ldots ,
1+\alpha _n-\beta _i; 1+\beta _1-\beta _i,{\hat{\ldots} },1+\beta _n-
\beta _i;z),
\ee
where $i=1,\ldots ,n$, $\beta _n=1$,  and the sign ${\hat{\ldots}} $ denotes
that the argument $1+\beta _i-\beta _i$ is omitted.

In [Po 1888] the following proposition  was proved.

\medskip\noindent
{\bf Lemma 2.} [Po 1888]. {\em If in} (24) $\gamma \not \in {\Bbb N}$
{\em then near the point} $z=1$ {\em equation} (23) {\em has} $n-1$
{\em analytic solutions of the form}
\be u_j(z)=(z-1)^{j-1}+O\left( (z-1)^{n-1}\right) \ee
{\em for } $j=1,\ldots ,n-1$ {\em corresponding to the exponents}
$0,1,\ldots ,n-2$.

Now we are ready to prove

\medskip\noindent{\bf Theorem 2.} {\em The function} $y(t)=x'(t)$
{\em has the following properties:}
\begin{description}
\item[1.] $y(t)$  {\em is analytic in the disc}
\be D_{r_n}=\{ t\,\colon |t|<r_n\},\qquad \mbox{where}\quad
r_n^{n-1}=\frac{(n-1)^{n-1}}{n^n}, \ee
{\em and the expansion of} $y(t)$ {\em as} $t\to 0$ {\em is of the form}
\be y(t)={\tilde y}(z)= 1+c_1z+c_2\,z^2+\cdots ,\qquad \mbox{where}\quad
z=t^{n-1}. \ee

\item[2.] {\em The function} ${\tilde y}(z)$ {\em is analytic on}
${\Bbb C}P^1$ {\em and its values are distinct from zero and infinity except
at points} $z=z_0=r_n^{n-1}=\gamma _n^{-1}$ {\em and} $z=\infty $.
{\em At} $z=z_0$ {\em it is equal to infinity, and at} $z=\infty $
{\em it vanishes.}

{\em As} $|t|\to \infty $, {\em we have }
\be x(t)=t\sum _{k=1}^{n-1} z^{-\alpha _k} g_k(z),\quad {\tilde y}(z)=
\sum _{k=1}^{n-1} z^{-\alpha _k}\,{\tilde g}_k(z),\quad \alpha _k=\frac{k}{n},
\quad z=t^{n-1}. \ee
{\em At} $z=\infty $, {\em the functions} $g_k(z)$ {\em and}
${\tilde g}_k(z)$ {\em are analytic and their values are distinct from zero
and infinity}.

\item[3.]
{\em Near the point} $z=z_0$, {\em the function} ${\tilde y}(z)$ {\em has
the form}
\be {\tilde y}(z)=(z-z_0)^{-1/2}f_1(z)+f_2(z), \ee
{\em where the functions} $f_1(z)$ {\em and} $f_2(z)$ {\em are analytic at}
$z=z_0$ {\em with} $f_1(z_0)\neq 0$ {\em and}  $f_2(z_0)\neq \infty $.

\item[4.] {\em The dimension of the linear space of all branches of the
function} ${\tilde y}(z)$ {\em obtained by going around the branch points}
$z=z_0$ {\em and} $z=\infty $ {\em is equal to} $n-1$.
\end{description}

\medskip\noindent{\em Proof}.  

\medskip\noindent 
{\bf 1.} The first statement of theorem follows from formulae (18), (19) and 
from the solution of equation (1) by iterations
\begin{eqnarray}
x &=& t +x^n,\\
x &=& t +(t+x^n)^n,\qquad x=t +\left( t+(t+x^n)^n\right) ^n,\ldots ,\\
x &=& t +\frac{c_1}{n}\,t^n+\frac{c_2}{2n-1}\,t^{2n-1}+\frac{c_3}{3n-2}\,
t^{3n-2}+\cdots ,\end{eqnarray}
\be c_1=n,\qquad c_2=\frac{(2n)(2n-1)}{1\cdot 2},\qquad c_3=\frac{3n\,(3n-1)
(3n-2)}{1\cdot 2\cdot 3},\cdots .\ee
The radius of convergence of this series is determined by the condition
that the discriminant of equation (1) vanishes. From this we derive
the formula
\be z_0=r^{n-1}_n=\gamma ^{-1}_n=\frac{(n-1)^{n-1}}{n^n}\,. \ee

\medskip\noindent
{\bf 2.} This follows from the iterations of formula
\be x(t)=\varepsilon _n\,t^{1/n}\left( 1-\frac{x}{t}\right) ^{1/n},\qquad
\mbox{where}\quad \varepsilon _n=\exp \left( i\,\frac{\pi }{n}\right), \qquad
\mbox{as}\quad |t|\to \infty . \ee

\medskip\noindent{\bf 3.}
In the same way, we obtain the expansion of ${\tilde y}(z)$ at the point
$z=z_0$.

\medskip\noindent{\bf 4.}
Equation (1) has $n$ solutions $x^{(j)}(t)$, $j=1,\ldots ,n$, $x^{(1)}(t)=
x(t)$, and we have $\sum x^{(j)}(t)=0$.

Therefore, the $n$ branches ${\tilde y}^{(1)}(z),\ldots ,{\tilde y}^{(n)}(z)$
of the algebraic function ${\tilde y}(z)=x'(t)$ are linearly dependent
\be {\tilde y}^{(1)}(z)+{\tilde y}^{(2)}(z)+\cdots +{\tilde y}^{(n)}(z)=0, \ee
and the dimension of the linear space of all branches of the function
${\tilde y}(z)$ is equal to $n-1$.

Now we are ready to prove Theorem 1.

We give two different proofs of this theorem. The first proof is based
on the general idea of a beautiful Riemann's paper [Rie 1857] adapted to
the problem under consideration. The second one uses the standard
calculation of the residue.

\newpage
\medskip\noindent{\em Proof 1.} From equations (1) and (2) it follows
that the function ${\tilde y}(z)=(1-nx^{n-1})^{-1}$, $z=t^{n-1}$, vanishes 
only at $z=\infty $, and it is equal to infinity only at $z=z_0$. Let 
${\tilde y}^{(1)}(z),\ldots ,{\tilde y}^{(n)}(z)$ be all the $n$ branches of
this function. Then
\be {\tilde y}^{(1)}(z)+{\tilde y}^{(2)}(z)+\cdots +{\tilde y}^{(n)}(z)=0, \ee
and the space $L_n$ of linear combinations of such functions is 
$(n-1)$-dimensional. At the points $z=0$, $z=z_0$ and $z=\infty $, 
the expansion of the function $f(z)\in L_n$ is defined by the exponents: 
\begin{eqnarray*}
\alpha _1=\frac1{n},\ldots ,\alpha _{n-1}=\frac{n-1}{n},&\qquad &
\mbox{at}\quad z=\infty ,\\
\beta _1=\frac1{n-1},\,\,\ldots ,\,\,\beta _{n-2}=
\frac{n-2}{n-1},&\qquad & \mbox{at}\quad z=0, \\
\gamma _1=0,\,\,\gamma _2=1,\ldots ,\gamma _{n-2}=n-3,\,\,\gamma _{n-1}=
-\,\frac12 &\qquad &\mbox{at}\quad z=z_0. \end{eqnarray*}
On the other hand, $n-1$ linearly independent solutions, $u_1(z),\ldots 
u_{n-1}(z)$, of the hypergeometric equation 
\be D\cdot (D+\beta _1-1)\cdots (D+\beta _{n-2}-1)\,u(z)-z(D+\alpha _1)\cdots 
(D+\alpha _{n-1})\,u(z)=0\ee
with $D=z\,{d}/{dz}$ have the same exponents and satisfy the conditions 
of the Riemann theorem. Therefore, according to this theorem 
\be {\tilde y}(z)={\tilde y}^{(1)}(z)\in L_n,\qquad {\tilde y}(z)= 
c_1\,u_1(z)+\cdots +c_{n-1}\,u_{n-1}(z). \ee 
The constants $c_k$  may be found from the consideration of the 
limit as $z\to 0$. As a result, we obtain formulae (20) and (21)\,\footnote
{\,Note that other branches of functions $y^{(j)}(t)$ and $x^{(j)}(t) 
\in L_n$, i.e., they are also the linear combinations of the functions 
$u_1(z),\ldots , u_{n-1}(z)$.}.

\medskip\noindent{\em Proof 2.} Here we use the calculation of the residue. 
We have 
\be {\tilde y}(z)=y_{n,0}(z)=1+c_1^{(0)}z+\cdots +c_k^{(0)}\,z^k+\cdots ,
\qquad z=t^{n-1}. \ee
Therefore,
\be \frac{c_k^{(0)}}{t}+\frac{df_k}{dt}=\frac{{\tilde y}(z)}
{x^{1+k(n-1)}(1-x^{(n-1)})^{1+k(n-1)}}, \ee
or
\be \frac{c_k^{(0)}}{t}+\frac{df_k}{dt}=\frac{C_k^{(0)}}{x}+\frac{dF_k}{dx}
\,.\ee
Here $f_k(t)$ and $F_k(x)$ are some rational functions.

The quantity $C_j^{(0)}$ is given by the expansion of the function
\be (1-z)^{-(1+k(n-1))}=\sum _{j=0}^{\infty } C_{j}^{(0)}z^j. \ee
Hence, 
\be c_k^{(0)}=C_k^{(0)}=\left( \begin{array}{c} k n \\
k\end{array} \right) . \ee 
The analogous calculation for the function 
\[ y_{n,j}(z)=t^j\,\sum _{k=0}^\infty c_k^{(j)}\,z^k \] 
gives
\[ c_k^{(j)}=\left( \begin{array}{c} kn+j\\ k \end{array} \right) . \]
This completes the proof.

From this it follows formula (7) for $y_{n,j}(t)$. Integrating 
$y_{n,j-1}(t)$ we obtain (21). 

So, Theorem 1 is proved.

\medskip\noindent{\bf 5.} We give here the additional properties of 
the functions $x_{n,j}$ and $y_{n,j}$. 

\medskip\noindent{\bf Lemma 3.}  {\em The function} $y(t)=y_{n,0}(t)$ 
{\em satisfies the following nonlinear differential equation} 
\be \frac1{y}\,\frac{d}{dt}\,\frac1{y}\,\frac{d}{dt}\,\cdots \frac1{y} 
=-\,n!\,. \ee 

\medskip\noindent{\em Proof.} We have 
\be \frac1{y}=1-nx^{n-1}. \ee Differentiating this $n-1$ times 
with respect to $t$, we obtain equation (46).

\medskip\noindent{\bf Lemma 4.} {\em For the function $y(t)$ there is 
the following formal integral representation}\,\footnote{\,\,The term 
{\em formal} means that we should first expand the function $\exp \left( 
{\bar z}\,(z+t)^n\right) $ in the power series and only then integrate it. 
Note that for the functions $x_{n,j}(t)$ and $y_{n,j}(t)$ we also have 
the standard integral representations (see [Po 1888], [BE 1953]).} 
\be y(t)=\int _{\Bbb C} d\mu (z,\bar z)\,\exp \left( {\bar z} (z+t)^n\right) ,
 \ee
{\em where} 
\be d\mu (z,{\bar z}) =\frac1{\pi }\,\exp \left( -|z|^2\right) dz\,
d{\bar z}\,. \ee 

\medskip\noindent{\em Proof.} 
We can check this by  expanding of the integrand in the series and using 
the formula 
\be \int d\mu (z,{\bar z})\,z^j\,{\bar z}^k=\delta _{jk}\,k!\,. \ee 

In conclusion, observe that from the evident formulae such as 
\be (x(t))^j(x(t))^l=(x(t))^{j+l},\qquad (x(t))^j(x(t)^l\,y(t))=(x(t))^{j+l}\,
y(t) \ee 
there follows nontrivial (and probably new) identities for the generalized 
hypergeometric functions. We give here a couple of them.

\medskip\noindent{\bf Lemma 5.} {\em The following identities are valid:} 
\begin{eqnarray}
&& _nF_{n-1}\left( \frac{j}{n},\ldots ,\frac{j+n-1}{n};\frac{j+1}{n-1},
\ldots ,\frac{j+n-1}{n-1};z\right) \nonumber \\
&\times & \,_nF_{n-1}\left( \frac{l}n,\ldots ,\frac{l+n-1}{n};\frac{l+1}{n-1},
\ldots ,\frac{l+n-1}{n-1};z\right) =\nonumber \\
&&\,_nF_{n-1}\left( \frac{j+l}n,\ldots ,\frac{j+l+n-1}n;\frac{j+l+1}{n-1},
\ldots ,\frac{j+l+n-1}{n-1};z\right) ,\nonumber \\
&&\end{eqnarray}
\begin{eqnarray}
&&\,_nF_{n-1}\left( \frac{j}n,\ldots ,\frac{j+n-1}n;\frac{j+1}{n-1},\ldots ,
\frac{j+n-1}{n-1};z\right) \nonumber \\
&\times &\,_{n-1}F_{n-2}\left(\frac{l+1}n,\ldots ,\frac{l+n}n;\frac{l+1}{n-1},
\ldots ,\frac{l+n-1}{n-1};z\right) =\nonumber \\
&&_{n-1}F_{n-2}\left( \frac{j+l}n,\ldots ,\frac{j+l+n}n;\frac{l+l+1}{n-1},
\ldots ,\frac{j+l+n-1}{n-1};z\right) .\nonumber \\
&&\end{eqnarray}

\newpage
\medskip\noindent
\section*{Appendix. Explicit expressions for functions
$x(t)=x_{n,1}(t)$ and $y_{j}(t)=y_{n,j}(t)$ for n =2, 3, 4, 5 and 6.}

\setcounter{equation}{0}

\medskip\noindent $n=2$
 
\begin{eqnarray*}
x(t) &=& t\,_2F_1\left( \frac12,1;2;4t\right) ,\\
x(t) &=& t+t^2+2\,t^3+5\,t^4+2\cdot 7\,t^5 +2\cdot 3\cdot 7\,t^6+
2^2\cdot 3\cdot 11\,t^7\\
&+& 3\cdot 11\cdot 13\, t^8+2\cdot 5\cdot 11\cdot 13\,t^9+\cdots
\end{eqnarray*}
\begin{eqnarray*}
y_0 &=& \,_1F_0\left( \frac12;4t\right) ,\\
y_0 &=& 1+2\,t+2\cdot 3\,t^2+2^2\cdot 5\,t^3\\
&+&2\cdot 5\cdot 7\,t^4 +2^2\cdot 3^2\cdot 7\,t^5+
2^2\cdot 3\cdot 7\cdot 11\,t^6 \\
&+& 2^3\cdot 3\cdot 11\cdot 13\,t^7+2\cdot 3^2\cdot 5\cdot 11
\cdot 13\,t^8+\cdots \\
y_1 &=& t+3\,t^2+2\cdot 5\,t^3+5\cdot 7\,t^4+
2\cdot 3^2\cdot 7\,t^5+\cdots
\end{eqnarray*}

\medskip\noindent $n=3$
\begin{eqnarray*}
x(t) &=& t\,_2F_1\left( \frac13,\frac23;\frac32;\frac{27}4t^2\right) \\
x(t) &=& t+t^3+3\,t^5+2^2\cdot 3\,t^7+5\cdot 11\,t^9 +
3\cdot 7\cdot 13\,t^{11}\\
&+& 2^2\cdot 3\cdot 7\cdot 17\,t^{13}+
2^3\cdot 3\cdot 17\cdot 19\,t^{15}+
3^2\cdot 11\cdot 19\cdot 23\,t^{17}+\cdots
\end{eqnarray*}
\begin{eqnarray*}
y_0(t) &=& \,_2F_1\left( \frac13,\frac23;\frac12;\frac{27}4t^2\right) ,\\
y_0(t)&=& y(t|3,0) =1+3\,t^2+3\cdot 5\,t^4+2^2\cdot 3\cdot 7\,t^6
+3^2\cdot 5\cdot 11\,t^8\\
&+&3\cdot 7\cdot 11\cdot 13\,t^{10}+
2^2\cdot 3\cdot 7\cdot 13\cdot 17\,t^{12}\\
&+&2^3\cdot 3^2\cdot 5\cdot 17\cdot 19\,t^{14}+
3^2\cdot 11\cdot 17\cdot 19\cdot 23\,t^{16}+\cdots \\
y_0^{-1}(t) &=& 1-3\,t^2-2\cdot 3\,t^4-3\cdot 7\,t^6-
2\cdot 3^2\cdot 5\,t^8-3\cdot 11\cdot 13\,t^{10}\\
&-& 2^3\cdot 3\cdot 7\cdot 13\,t^{12} -\cdots
\end{eqnarray*}
\begin{eqnarray*}
y_1(t)&=&t\,_2F_1\left( \frac23, \frac43; \frac32; \frac{27}4t^2\right) \\
y_1(t)&=& t+2^2\,t^3+3\cdot 7\,t^5+2^3\cdot 3\cdot 5
\,t^7+5\cdot 11\cdot 13\,t^9 \\
&+& 2^4\cdot 3\cdot 7\cdot 13\,t^{11}+
2^2\cdot 3\cdot 7\cdot 17\cdot 19\,t^{13}+\cdots
\end{eqnarray*}
\begin{eqnarray*}
y_2(t) &=& t^2\,_3F_2\left( 1,\frac43, \frac53; \frac32, 2;
\frac{3^3}{2^2}t^2\right) \\
y_2(t) &=& t^2+5\,t^4+2^2\cdot 7\,t^6 +3\cdot 5\cdot 11\,t^8\\
&+&7\cdot 11\cdot 13\,t^{10}+\cdots
\end{eqnarray*}

\medskip\noindent $n=4$
\begin{eqnarray*}
x(t) &=& t\,_3F_2\left( \frac14,\frac12,\frac34;\frac23,\frac43;\frac{4^4}
{3^3}t^3\right) \\
x(t) &=& t+t^4+2^2\,t^7+2\cdot 11\,t^{10}+
2^2\cdot 5\cdot 7\,t^{13}\\
&+& 3\cdot 17\cdot 19\,t^{16} +
2^2\cdot 7\cdot 11\cdot 23\,t^{19}+
2^2\cdot 3^2\cdot 5\cdot 13\cdot 23\,t^{22}\\
&+& 2^2\cdot 3^2\cdot 13\cdot 29\cdot 31\,t^{25}+\cdots
\end{eqnarray*}
\begin{eqnarray*}
y_0(t)&=& \,_3F_2\left( \frac14, \frac12, \frac34; \frac13,\frac 23;
\frac{4^4}{3^3}t^3\right) \\
y_0(t) &=&  1+2^2\,t^3+2^2\cdot 7\,t^6+
2^2\cdot 5\cdot 11\,t^9+2^2\cdot 5\cdot 7\cdot 13\,t^{12} \\
&+& 2^4\cdot 3\cdot 17\cdot 19\,t^{15}+
2^2\cdot 7\cdot 11\cdot 19\cdot 23\,t^{18}\\
&+& 2^3\cdot 3^2\cdot 5\cdot 11\cdot 13\cdot 23\,t^{21}+
2^2\cdot 3^2\cdot 5^2\cdot 13\cdot 29\cdot 31\,t^{24}+\cdots
\end{eqnarray*}
\begin{eqnarray*}
y_1(t) &=& t\,_3F_2\left( \frac12,\frac34, \frac54; \frac23,
\frac43; \frac{4^4}{3^3}t^3\right) \\
y_1(t) &=& t+5\,t^4+2^2\cdot 3^2\,t^7+2\cdot 11\cdot 13\,t^{10}\\
&+& 2^2\cdot 5\cdot 7\cdot 17\,t^{13} +\cdots
\end{eqnarray*}
\begin{eqnarray*}
y_2(t) &=& t^2\,_3F_2\left( \frac34, \frac54, \frac32; \frac43,
\frac53; \frac{4^4}{3^3}t^3\right) \\
y_2(t) &=& t^2+2\cdot 3\,t^5+3^2\cdot 5\,t^8+
2^2\cdot 7\cdot 13\,t^{11}\\
&+& 2^2\cdot 3^2\cdot 5\cdot 17\,t^{14}+\cdots
\end{eqnarray*}
\begin{eqnarray*}
y_3(t) &=& t^3\,_4F_3\left( 1, \frac54, \frac32, \frac74;
\frac43, \frac53, 2; \frac{4^4}{3^3}t^3\right) \\
y_3(t) &=& t^3+7\,t^6+5\cdot 11\,t^9 +5\cdot 7\cdot 13\,t^{12}\\
&+& 2^2\cdot 3\cdot 17\cdot 19\,t^{15} +\cdots
\end{eqnarray*}

\medskip\noindent $n=5$
\begin{eqnarray*}
x(t) &=& t\,_4F_3\left( \frac15,\frac25,\frac35,\frac45;\frac12,\frac34,
\frac54;\frac{5^5}{4^4}t^4\right) \\
x(t) &=& t+t^5+5\,t^9+5\cdot 7\,t^{13}+3\cdot 5\cdot 19\,t^{17}+
2\cdot 5\cdot 11\cdot 23\,t^{21}\\
&+& 3^2\cdot 7\cdot 13\cdot 29\,t^{25}+
2^3\cdot 5\cdot 11\cdot 17\cdot 31\,t^{29}\\
&+&3\cdot 5\cdot 13\cdot 17\cdot 19\cdot 37\,t^{33}+\cdots
\end{eqnarray*}
\begin{eqnarray*}
y_0(t) &=& \,_4F_3\left( \frac15, \frac25, \frac35, \frac45; \frac14,
\frac12, \frac34; \frac{5^5}{4^4}t^4 \right) \\
y_0(t) &=& 1+5\,t^4+3^2\cdot 5\,t^8+
5\cdot 7\cdot 13\,t^{12}+3\cdot 5\cdot 17\cdot 19\,t^{16}\\
&+&2\cdot 3\cdot 5\cdot 7\cdot 11\cdot 23\,t^{20}+
3^2\cdot 5^2\cdot 7\cdot 13\cdot 29\,t^{24}\\
&+& 2^3\cdot 5\cdot 11\cdot 17\cdot 29\cdot 31\,t^{28}\\
&+&3^2\cdot 5\cdot 11\cdot 13\cdot 17\cdot 19\cdot 37\,t^{32}+
\cdots
\end{eqnarray*}
\begin{eqnarray*}
y_1(t) &=& t\,_4F_3\left( \frac25, \frac35, \frac45, \frac65;
\frac12,\frac34, \frac54;\frac{5^5}{4^4}t^4\right) \\
y_1(t) &=& t+2\cdot 3\,t^5+5\cdot 11\,t^9+
2^4\cdot 5\cdot 7\,t^{13} \\
&+& 3^2\cdot 5\cdot 7\cdot 19\,t^{17}+\cdots
\end{eqnarray*}
\begin{eqnarray*}
y_2(t) &=& t^2\,_4F_3\left( \frac35,\frac45, \frac65,\frac75;
\frac34, \frac54,\frac32; \frac{5^5}{4^4}t^4\right) \\
y_2(t) &=& t^2+7\,t^6+2\cdot 3\cdot 11\,t^{10}+
2^3\cdot 5\cdot 17\,t^{14} \\
&+& 5\cdot 7\cdot 11\cdot 19\,t^{18}+
2\cdot 3^3\cdot 5\cdot 13\cdot 23\,t^{22}+\cdots
\end{eqnarray*}
\begin{eqnarray*}
y_3(t) &=& t^3\,_4F_3\left( \frac45, \frac65, \frac75,\frac85; 
\frac54, \frac32, \frac74; \frac{5^5}{4^4}t^4\right) \\
y_3(t) &=& t^3+2^3\,t^7+2\cdot 3\cdot 13\,t^{11} +
2^4\cdot 3\cdot 17\,t^{15}\\
&+& 5\cdot 7\cdot 11\cdot 23\,t^{19}+
2^3\cdot 3^3\cdot 5\cdot 7\cdot 13\,t^{23}+\cdots
\end{eqnarray*}
\begin{eqnarray*}
y_4(t) &=& t^4\,_5F_4\left( 1,\frac65,\frac75,\frac85, \frac95;
\frac54, \frac32, \frac74;2; \frac{5^5}{4^4}t^4\right) \\
y_4(t) &=& t^4+3^2\,t^8+7\cdot 13\,t^{12}+
3\cdot 17\cdot 19\,t^{16}\\
&+& 2\cdot 3\cdot 7\cdot 11\cdot 23\,t^{20}
+3^2\cdot 5\cdot 7\cdot 13\cdot 29\,t^{24}+\cdots
\end{eqnarray*}

\medskip\noindent$n=6$
\begin{eqnarray*}
x(t) &=& t\,_5F_4\left( \frac16, \frac13, \frac12, \frac23, \frac56;
\frac25, \frac35, \frac45, \frac65; \frac{6^6}{5^5}t^5\right) \\
x(t) &=& t+t^6+2\cdot 3\,t^{11}+3\cdot 17\,t^{16}+
2\cdot 11\cdot 23\,t^{21}\\
&+& 3^3\cdot 7\cdot 29\,t^{26}+
2^4\cdot 3\cdot 7\cdot 11\cdot 17\,t^{31}\\
&+& 2\cdot 13\cdot 19\cdot 37\cdot 41\,t^{36}+
2\cdot 3^2\cdot 11\cdot 23\cdot 43\cdot 47\,t^{41}+\cdots
\end{eqnarray*}
\begin{eqnarray*}
y_0(t) &=& _5F_4\left( \frac16, \frac13, \frac12, \frac23, \frac56;
\frac15, \frac25, \frac35, \frac45; \frac{6^6}{5^5}t^6\right) \\
y_0(t) &=& 1+2\cdot 3\,t^5+2\cdot 3\cdot 11\,t^{10}+
2^4\cdot 3\cdot 17\,t^{15}\\
&+&2\cdot 3\cdot 7\cdot 11\cdot 23\,t^{20}+
2\cdot 3^3\cdot 7\cdot 13\cdot 29\,t^{25}\\
&+& 2^4\cdot 3\cdot 7\cdot 11\cdot 17\cdot 31\,t^{30}+
2^3\cdot 3^2\cdot 13\cdot 19\cdot 37\cdot 41\,t^{35}\\
&+& 2\cdot 3^2\cdot 11\cdot 23\cdot 41\cdot 43\cdot 47\,t^{40}
+\cdots
\end{eqnarray*}

\newpage
\medskip\noindent{\bf Acknowledgments.} I am grateful to Professors
F. Hirzebruch, D. Leites, A. Levine and Yu.I. Manin for useful remarks
and the Max-Planck-Institute f\"ur Mathematik, Bonn for hospitality.

\end{document}